\documentclass[pdflatex,sn-basic]{sn-jnl}

\usepackage{graphicx} 

\def\paragraph#1{\noindent\textbf{#1} } 
\usepackage{natbib}
\usepackage{color}
\usepackage{caption}
\usepackage[title]{appendix}
\usepackage{graphicx}%
\usepackage{multirow}%
\usepackage{amsmath,amssymb,amsfonts}%
\usepackage{amsthm}%
\usepackage{mathrsfs}%
\usepackage[title]{appendix}%
\usepackage{xcolor}%
\usepackage{textcomp}%
\usepackage{manyfoot}%
\usepackage{booktabs}%
\usepackage{algorithm}%
\usepackage{algorithmicx}%
\usepackage{algpseudocode}%
\usepackage{listings}%
\usepackage{comment}%
\newif\ifwithauthors
\withauthorstrue   

\begin{document}

\title{City Models: Past, Present and Future Prospects}



\ifwithauthors

\author[2]{\fnm{Helge} \sur{Ritter}}
\author[1,4]{\fnm{Otthein} \sur{Herzog*}}
\author[3]{\fnm{Kurt} \sur{Rothermel}}
\author[5,6]{\fnm{Anthony G.} \sur{Cohn}}
\author[1]{\fnm{Zhiqiang} \sur{Wu}}

\affil[1]{\orgdiv{College of Architecture and Urban Planning}, \orgname{Tongji University}, 

\orgaddress{ \street{1239 Siping Road}, \city{Shanghai}  \postcode{200092}, \country{PRC}}}

\affil[2]{\orgdiv{Faculty of Technology}, \orgname{Bielefeld University}, \orgaddress{\street{Universitätsstr.25},

\postcode{33615} \city{Bielefeld}, \country{Germany}}}

\affil[3]{\orgdiv{Institute of Parallel and Distributed Systems}, \orgname{University of Stuttgart}, \orgaddress{\street{Universitätsstr. 38}, \postcode{70569} \city{Stuttgart}, \country{Germany}}}

\affil[4]{\orgdiv{FB 03}, \orgname{University of Bremen}, \orgaddress{\street{Bibliotheksstr. 5}, 

\postcode{28539} \city{Bremen},  \country{Germany}}}

\affil[5]{\orgdiv{School of Computer Science}, \orgname{University of Leeds}, \city{Leeds}, \postcode{LS2 9JT}, \country{United Kingdom}}

\affil[6]{\orgname{The Alan Turing Institute}, \city{London}, \postcode{ NW1 2DB}, \country{United Kingdom}}


\maketitle

\begin{center}
Date 2025-03-10
\end{center}







\newpage

\else
\maketitle
\newpage
\fi

\newcommand{\tcg}[1]{#1}

\section{Abstract}
This paper attempts to take a comprehensive 
look at the challenges of representing the spatio-temporal structures and dynamic processes that
define a city's overall characteristics. For the task of urban planning and urban operation, we take the stance that even if the necessary representations of these structures and processes can be achieved, the most important representation of the relevant mindsets of the citizens are, unfortunately, mostly neglected. 

After a  review of major ``traditional" urban models of structures behind urban scale, form, and dynamics, we turn
to major recent modeling approaches triggered by 
recent advances in AI that enable multimodal generative models. Some of these models can create
representations of geometries, networks and images, and reason
flexibly at a human-compatible semantic level. They provide huge amounts of knowledge
extracted from huge collections of text and image documents and cover the required rich representation spectrum including geographic knowledge by different knowledge sources, degrees of granularity and scales.

We then discuss what these new opportunities mean for coping
with the modeling challenges posed by cities, in particular with
regard to the role and impact of citizens and their interactions
within the city infrastructure. 
We propose to integrate these possibilities with existing approaches, such as agent-based models, which opens up new modeling
spaces including rich citizen models which are able to also represent  social interactions.

Finally, we  put forward some thoughts
about a vision of a ``social AI in a city ecosystem'' that adds relevant citizen models to state-of-the-art structural and process models. This extended city representation will enable urban planners to establish citizen-oriented planning of city infrastructures, to make them into inviting environments
that reconcile and foster human culture, city resilience and sustainability.

\section{Introduction: Modeling Cities -- a multifaceted challenge}

The complexity of cities makes their modeling into a multifaceted challenge,
arising from the numerous perspectives to be covered by models.

\begin{figure}
    \centering
    
    \includegraphics[width=0.48\columnwidth]{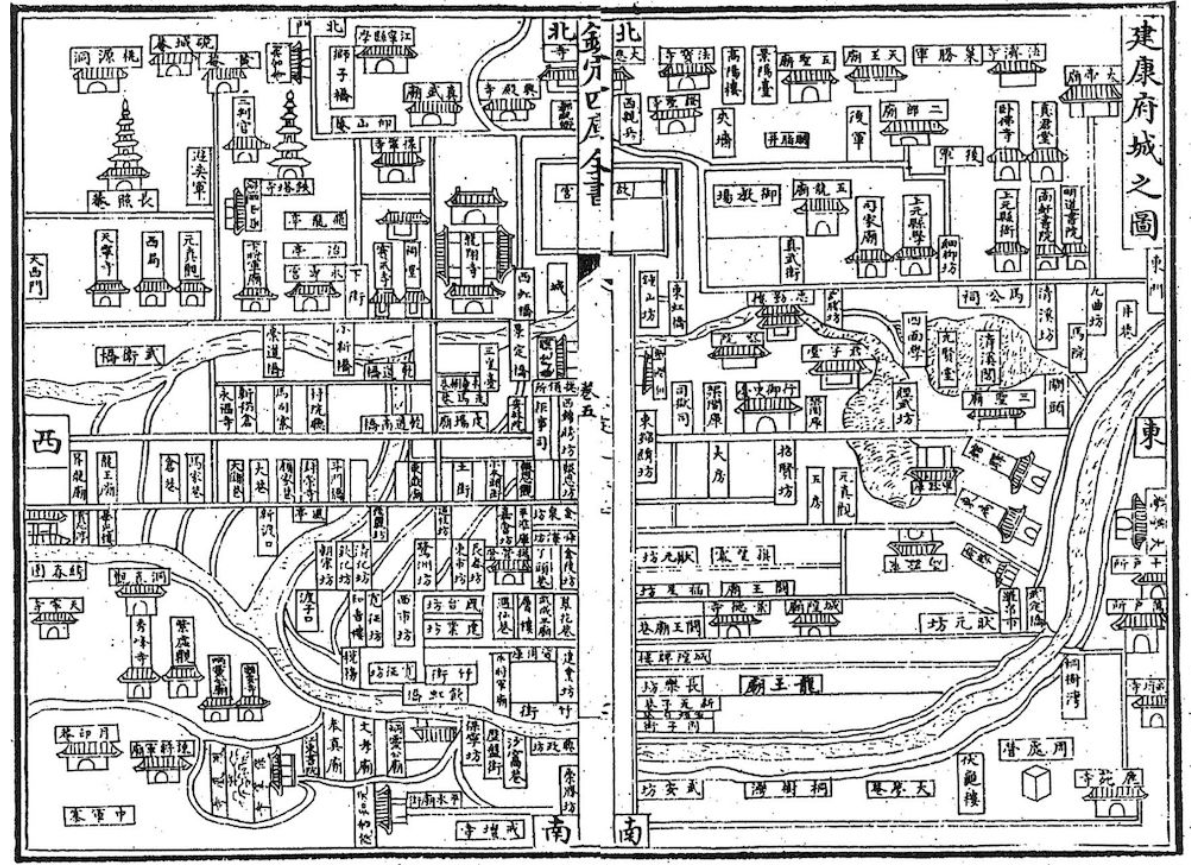}\hspace*{3mm}\includegraphics[width=0.5\columnwidth]{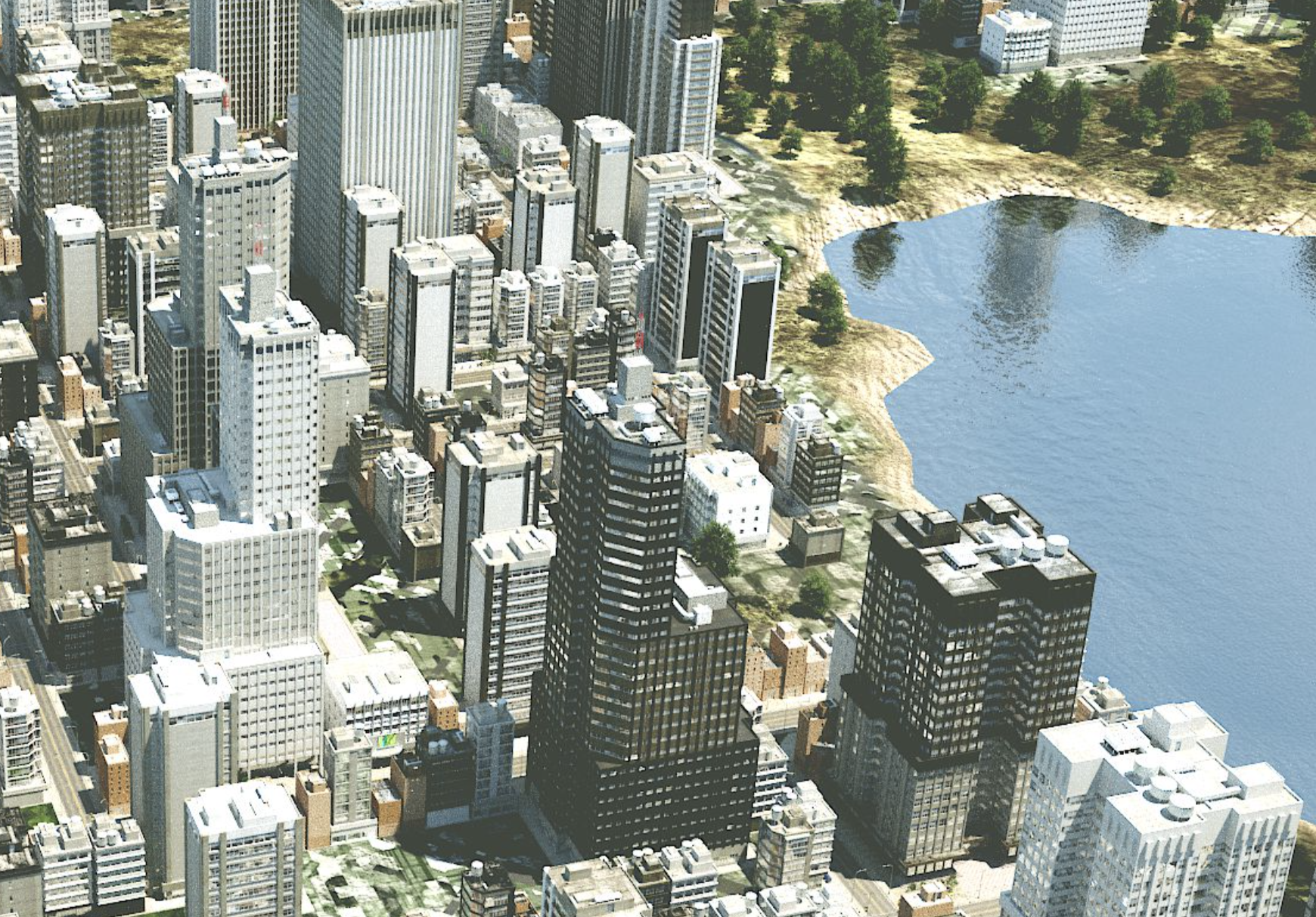}
    \caption{\textit{Left:} Ancient printed map of the city of Nanjing (around 12th century \citep{nanjingmap}). \textit{Right:} 3D model of part of a city model, procedurally generated with Blender \citep{blender} and the CitiGen software plugin \citep{citygen}.}
    \label{fig:citymodels_old_and_new}
\end{figure}

Historically, the perhaps earliest city models started with pictures or crude maps
to aid finding one's way through the city. With increasing size and complexity of
cities, along with improved technical means, city maps improved steadily, with
the first printed maps of major cities such as Rome, Antwerp, and Paris appearing
in Europe between the 15th and 16th centuries, with printed city maps in China dating back at least to the 12th century,
with maps carved in stone even older \citep{cheng2022nonscientific}. These may be considered as the earliest
prototypes of spatial city models that already also cared about visualization of
and semantic information about prominent landmarks, addressing in their essence very similar questions as the much more
detailed city models of today (Fig \ref{fig:citymodels_old_and_new}).

Modern cartography had its beginnings around the 18th century, followed by major
advances, such as triangulation networks and aerial image capture in the 19th century,
and finally the availability of digital computers in the 20th century, enabling geospatial
information systems and the modern digital city maps of our time. The addition
of modern Building Information Modeling (BIM) systems towards the turn of the millennium
extended these 2D models into fully three-dimensional city models.

The early city models were essentially \textit{static} and
focused on \textit{modeling shape} (but maps of other features,
such as weather maps, or of the spatial distribution of a wide range of entities, including
population, did exist as well). However, these models were purely descriptive and
always focused on a single particular instance (such as a map of Rome). However, visualization triggers for more flexibility, e.g., to realize different views with regard to perspective, scale, illumination, weather conditions or seasons in a year. All this requires a model that can render or \textit{simulate a variety of views} under different conditions. Today,
such view simulation can happen instantly,
including simulated changes to the geometry or even presence of entire buildings or other
structures. We owe such models to our reasonably accurate knowledge of the physical
laws that are involved when surfaces reflect light to form a picture for an
observer, along with powerful computing hardware to simulate the imaging process.
We also may wish
models that reflect other properties of cities, such as traffic flows, resilience
against flooding or heat, pedestrian flows, and many others. Such models help us \textit{explore consequences changing the city itself}, for instance, by generating alternative plausible city variants ``of a certain kind" (e.g., a mid-sized
town in a particular region with a particular mix of building styles for its houses). This includes  \textit{making predictions}, e.g., generating probable development paths of a given (true or hypothetical) city in the future. Finally, modeling variability can help us synthesize developmental patterns of an entire range of
cities that share selected similarities at present. 

All these modeling directions differ from our initial visualization example by involving processes for which we do not
know good law-like representations to implement a simulation.
This problem becomes even more acute when we turn to the impact of \textit{the citizens}
since the actions of the citizens are a major, if not the strongest, driver of temporal change.
Citizens act in very complex and often highly interdependent ways, giving rise to complex patterns of change at many different time scales, from hours to decades. Although some aspects of their
actions may admit aggregated models based on average behavior, there are always
macroscopic changes in a city that have originated from decisions of only few or even single individuals: the physical infrastructure is much simpler than the main drivers of its dynamics, the citizens. 

This paper will first offer (Section 2) a view on the landscape of existing models, along their path from largely static geometry models to dynamic models, with an emphasis on major ideas, such as fractal geometry, cellular automata, dynamical systems, digital twins and multi-agent models, and how these attempts
led to accounting
for the important role of citizens. Then it will shift the focus to generative models and recent advances of AI that have enabled their construction in a purely data-driven way (Section 3). Our 
emphasis
will be on the capabilities of models that can generate complex data objects, such as images or other spatial structures, and, particularly, on Large Language Models (LLMs). These models can learn, capture huge amounts of knowledge and bring it to bear on solving tasks that can be specified in natural language, or even in multimodal ways that admit combinations with images and sound. Finally, we will look at the new opportunities and prospects (Section 3) that arise from these innovations and present some thoughts on a vision of a ``social AI in city ecosystems'' that will enable urban planners to establish citizen-oriented planning of city infrastructures and to make them into inviting environments
that reconcile and foster human culture, city resilience and sustainability.

\section{The landscape of urban models: scale, form and dynamics}

The complexity of cities inevitably calls for many models, each one
addressing a subset of features that contribute to characterizing
a city. These models will come in many variants, but they all somehow have
to say something about how cities vary with scale, in spatial form, and how they vary in time.
Moreover, what they have to say can be represented at different levels
of abstraction or granularity. 

\paragraph{Role of scales.}
Cities and their structures span across many scales, both in the
spatial (meters to kilometers) as well as in the time dimension (minutes to hours, days and months). 

Spatial and temporal scales are linked in many ways, sometimes following simple patterns, such as changes to large structures usually taking more time to happen.

Complex couplings arise through the spatial coupling of moving
entities, such as pedestrians and cars. These couplings and the time
scales of their dynamics can vary abruptly when certain ``order parameters",
such as density, or cross critical thresholds are met \citep{helbing2001traffic}. 
Such ``emergent" or ``critical phenomena" can make it hard or
impossible to confine a model to a single time scale and instead
one must
strive for models that can capture such critical phenomena at the
``edge of chaos" \citep{chen2009urban}. Moreover, these conditions are not 
exceptional but a frequent characteristic of complex systems
\citep{siegenfeld2020introduction} and actually prerequisites for important capabilities, such
as resilience and innovation. 
Many elements of a city have memory-like properties, which
range from hysteresis over simple adaptation to sophisticated
planning and anticipation, which can act in complex, stabilizing or de-stabilizing
ways, e.g., when the decision behavior of citizens comes
into play, and extend to very long time scales, e.g., due to citizens' ability to create and pursue long-term plans, or slow processes such as demographic change.
                  
\paragraph{Scaling laws} express important aspects of scale. With regard to cities, properties such as area, energy consumption, or various connectivity measures (e.g., total road length), necessarily vary as a function of city size (e.g., measured by the number $p$ of inhabitants). It turns out that this variation often takes the form of a \textit{power law} for the quantity of interest:
$
   A = p^{d},
$
where for many cases, the expected values for the exponents $d$ deviate from their ``naive" expectations. These deviations often indicate
\textit{efficiency gains}, e.g., sublinear ($d<1$) growth behavior for quantities such as city area or total road length \citep{molinero2021geometry,chen2021scaling}. The resulting, usually non-integral exponents expose deviations from familiar, 1D or 2D geometries of road networks or occupied city area, and indicate linkages between scaling laws and models of \textit{urban form} (see below).

\paragraph{Models of urban form.}
Maps primarily embrace
the \textit{geometric aspects} of a city, in the
form of 2D, 3D or fractal models. 
Models
can be further 
distinguished by their focus on \textit{land usage} (2D areas), \textit{connectivity}
(graph-based representations), \textit{buildings} (3D structures) or combinations thereof (e.g.,
planar maps with street networks and embedded areas). These 

models
differ in their extent, e.g., buildings
are very localized whereas streets or other transportation facilities
can extend across wide ranges. 

Statistical models can
simplify local detail below a particular scale into averaged local bulk 
properties, e.g., of city patches. This can make shape properties, such as, e.g.,
distance or area, scale dependent. Fractal models  (see below)
offer an elegant solution to represent such nested-ness of detail.

\paragraph{Interaction and Proximity.}
Interactions between citizens are crucial
for any city and are usually sensitive to proximity. Land features such as ridges, rivers or lakes, or man-made structures such as fences and building complexes can impact on travel times, effectively ``distorting" geometric distance \citep{barnsley2014fractals}).
Urban road and transportation networks add further 
reshaping effects on spatio-temporal characteristics of interactions.
The resulting complex intertwining of many factors leads to significant
deviations of proximity from what would be expected according to Euclidean geometry. 

\paragraph{Graph- and Network-Models.} 
With the advent of wireless networks, which complement physical proximity 
with a completely new layer of ``virtual" proximity, network and graph models,
cities have become augmented with an extremely important
new layer of transport structures that needs
to be included in urban modeling.

Graph and network models offer a unified approach to model both the physical as well
as the information transport networks in this layer as
general connectivity structures, with
geometric analogs to distance, connectedness, and centrality. Because of their discretized nature, graphs are also very suitable
to represent  \textit{topological properties}, such as connectivity, which are
more coarse-grained than geometry. Graphs with weighted edges can also
conveniently represent travel distances or times 
instead of Euclidean distance \citep{sun2012fractal}. Using such models can
avoid overfitting while still representing important properties, such
as reachability or clusterings.

\paragraph{Fractal models}
permit a unified treatment of urban form, proximity structures and scaling properties \citep{batty1986fractal,batty1996preliminary,chen2021scaling}.
Moreover, these models have been observed to also capture intriguing
aspects of the \textit{growth of cities}
\citep{batty1989urban,shen2002fractal,zhang2022fractal}.
Fractal geometry
can be characterized in several ways,
for example by 
\textit{fractal dimensions} which describe how some size measure S (such
as length, area, volume) associated with a geometric object behaves under
a linear change of scale. This is described by a power law
$S(L)\propto L^D$ where L denotes the scaling factor and D$>$0 is the dimension
that characterizes the geometry of the object. For fractal objects, this
dimension can deviate from the usual integer values.

Besides area, other size measures, such
as the total length of road or transportation networks, or the length
of coast lines, also exhibit fractal scaling properties indicated by
non-integer D dimensions. As a result, the concept of fractal geometry
has been found to be useful for characterizing the scaling properties of
a variety of urban structures \citep{chen2021scaling,jahanmiri2022overview}.

\paragraph{Self-similarity and procedural city models.}
Fractal scaling behavior is closely linked to the property of \textit{self-similarity} -- 
i.e., when parts of a shape resemble a scaled version of itself or some part of it.
This can be turned around, creating a shape from suitably (either
deterministically or stochastically) re-arranged copies of a number
of initial ``generators'', and repeating this procedure recursively.
Such constructions can be shown to lead to geometries with
fractal dimensions that can be directly computed from their recursive
construction process. They turn out to offer approximations to many shapes
that are encountered in nature \citep{barnsley2014fractals,mandelbrot1983fractal}, as well as 
a number of man-made artefacts (buildings, cities, road networks).
There are different mathematical frameworks to formalize the generative
process, such as Lindenmayer- or L-Systems \citep{prusinkiewicz1986graphical} or iterated function systems \citep{barnsley2014fractals}.
They triggered the seminal work of \citet{parish2001procedural} which led to
increasingly refined \textit{procedural city
models} \citep{kelly2006survey,vanegas2009modelling,kim2018procedural},
and sophisticated city modeling engines, such as CityEngine
\citep{roumpani2022procedural} which has become widely available in the
commercial ESRI software \citep{cityengine2024}. Starting from a given
geographical area with land features and additional constraints, such as
targeted population density, these models can be parametrized to generate
house, land use, and street patterns that cover
the given area with an arrangement of houses and streets that follows
the prescribed constraints. In a subsequent step, the 2D geometry plan can be augmented
with further features, such as building heights and texture choices
for walls and roofs (an illustrative result is depicted in Fig. \ref{fig:citymodels_old_and_new} (right)).

\paragraph{Qualitative Spatial Representations (QSR)}
\citep{cohn2008qualitative,chen2015survey}
is a well developed field which is concerned with the representation of qualitative spatial information and reasoning with it.  In natural language, spatial information is usually represented qualitatively  (using prepositions such as \textit{on, in, left of, part of, under, touching, east of}, ...) and many  formalisms have been developed to represent such information. Each of these calculi consists of a set of small finite set of relations  -- e.g., one well known calculus (RCC)  for handling topological information consists of eight \textit{jointly exhaustive and pairwise disjoint} (JEPD) \textit{base} relations allowing one to specify whether two regions are disconnected, touching, partially overlapping, equal, or part of the other one (either touching at the boundary or not); RCC has been implemented in GeoSPARQL. Other calculi  focus on representing information about shape, size, directions and may also explicitly handle the case of regions with vague boundaries which is a common occurrence in geographic situations (e.g., exactly where is ``downtown''?).
Each calculus has associated  specialized inference mechanisms, in particular an efficient way to compute the \emph{composition} of two relations -- e.g., if one knows that region x is part of region y and region $y$ is part of region $z$, then one can infer that $x$ must be part of $z$.  
Using composition one can not only infer relationships not explicitly in a knowledge base, but also check if a database of facts concerning a set of spatial entities is consistent or not -- this might not be the case if the information has been provided by different agents or sensors, or at different times, for example.

\paragraph{Spatial economics.} Another class of models employs metric spaces with
a continuous distance metric to represent interactions
in cities with partial differential equations, similar to field equations
in physics. Taking further guidance from the idea of \textit{gravitation} choices for the placement of cities or economic
entities are modeled as if seeking minima of potential energy in a fictitious, gravitation-like field 
of attractive forces between entities for which proximity has an economically favorable effect.
This has given rise to spatial models of urban economics
that offer an attractive semi-analytic middle ground between highly abstract 
conceptual and extremely detailed simulation models \citep{redding2017quantitative}. 

\paragraph{Other spatial models.} Spatial cluster structures are  the
focus of  
the \textit{space syntax approach} to city modeling \citep{Hillier1976Space,van2021introduction}
which is a family of techniques for representing and analysing spatial relationships, especially in an urban context -- it views cities as  networks of spatial entities formed by the placing, grouping and orientation of buildings, enabling analyses of how streets interact spatially with all the other streets in a particular urban setting. It can be used to understand and predict the effects of spatial configurations on social, economic, and environmental behaviors within urban environments. By examining how streets, buildings, and public spaces are interconnected, space syntax provides insights into the movement patterns, accessibility, and spatial usage within a city. It thus has the potential to help urban planners and architects design more efficient, navigable, and sustainable urban environments by revealing the underlying spatial structure that influences human behaviour. A historically prominent example is \textit{central place theory} \citet{Chen_2006} that attempts to capture urban structure in a hierarchical set of \textit{central places} and their neighborhood
relationships.

\paragraph{City Development and Dynamics} 
has been a long-standing concern of urban researchers.
\textit{Diffusion-limited aggregation (DLA)} \citep{witten1983diffusion,halsey2000diffusion} is an early example of a model 
that takes the degree of simplification to the extreme. It models a growth dynamics
arising from a large number of particles moving along the random paths
of Brownian motion.

This dynamics causes an initial seed to grow into a dendritic pattern that 
can mimic the development of fractal city shapes over time \citep{fotheringham1989diffusion,batty1989urban}.
This has triggered 
numerous studies of 
self-organized models for growth processes in nature for their capability to also mimic dynamical
processes observed in the development of urban structures \citep{xu2007spatiotemporal,xu2021Emergence}.

\paragraph{\textbf{Early dynamical city models.}} Dynamical city modeling
originated with the classical Urban Dynamics model  \citep{forrester1970urban,moody1970urban}.
It uses a parameterized urban space of fixed size with a population
which distinguishes managers, skilled and unskilled workers. These
seek dwellings in three types of city areas and get their wages from
a number of firms categorized into new, mature and old firms, reflecting
the different needs of the different worker types.
The model is not a spatial model in that it uses aggregated quantities
for which it sets up rate equations for their temporal changes.
The rate of these changes is specified with equations
inspired by classical engineering control systems,
emphasizing the important role of closed feedback loops.
Although a major advance at its time, 
many of the limitations of this model were already noted by Forrester himself, amongst the most severe perhaps that
the city cannot expand and that the action response patterns of the agents
are taken as constant over an entire multi-year
simulation time span.

\paragraph{Cellular Automata (CA) models} 
provide an elegant
and intuitive approach to connect dynamics with spatial structure.
Popularized with Conway's ``Game of Life", \citet{tobler1979cellular} then
brought the framework to geographic modeling.
Space is modeled
as a grid (or a more general type of graph) of cells. Each cell holds a ``state variable''
describing local properties (such as type of land use, population density etc.) of the
corresponding land patch at a particular point in time. A ``transition rule'' then specifies, for each cell, its (possibly altered) state at the successive time step, using only the current state of the cell, and the states of cells within a certain spatial neighborhood.
Applying this rule to all cells iteratively creates the dynamics of the system. 
The remarkable complexity of the spatio-temporal patterns that can arise even from seemingly simple rules and neighborhoods
have led to a wide adoption of CA models for geographic and urban modeling \citep{white2000high,isinkaralar2023cellular}. Identifying the rule application with decisions of some ``agent" connects CA models with the wider class of agent based models.

A famous example is the \textit{Schelling model} and its subsequent variants
\citep{jensen2022introducing}. The model consists of two populations of agents
(e.g., ``white'' and ``black'') that make repeated decisions on whether to stay in
their current location or move to a different vacant place in the grid, with this decision only  dependent on the agent distribution in their neighborhood. When the considered alternative place has a more preferable (as defined by a utility function) neighborhood, they move; otherwise they stay. 
Simulating this
model has consistently shown that the model converges to a self-organized demixing of the
population into homogeneous domains. Subsequent studies 
allow  for more realism (such as including
market mechanisms for house prices, non-symmetric preference functions,
bounded rationality of agents, and generalization to more than two-component ethnic 
mixtures) have reproduced this type of behavior for a wide range of settings \citep{clark2008understanding}.

\paragraph{Dynamical Models of transportation systems}
 address another key urban structure.
Travel demand models aim to predict the response of transportation demand to modifications of the parameters of the transportation system and the 
travelers \citep{Bhat1999}.
Trip-based travel demand models, which were developed in the late 1950's, used individual trips as the unit of analysis. They assumed that it was possible to predict travel behavior based on household data averaged over a zone. Activity-based travel demand models developed in the late 1970s understand travel as a derived demand from the need to pursue activity distributed in space \citep{Axhausen1992} and hence focus on household and personal activity scheduling rather than trips isolated from human behavior. 
In recent years, deep learning approaches have been proposed for both short- and long-term  traffic forecasting \citep{Yin2022,Jiang2022}. These data-driven methods mine the complex non-linear spatial-temporal patterns of traffic data to make traffic predictions.
However, such data is only collected  infrequently and expensively from traffic surveys resulting in out-of-date predictions, although 
call records from cellular communication networks may provide an alternative \citep{Yin2018, Wang2024}.

\paragraph{Agent-based models} employ mobile agents located in space in order
to arrive at finer-grained models that can address spatio-temporal urban and land use phenomena \citep{huang2014review,zhang2021agent,crooks2021agent}.
The
simulated agents 
are endowed with some form of
behavior that has to be specified, ranging from extremely simple to highly sophisticated and
only limited by the computational resources of the simulator and the imagination/creativity of the modeler.

Typically, these agent-based models exploit Cellular Automata, where agents can change between
a small number of states and are fixed or limited to simple moves on some discrete
grid. Their state changes follow a set of rules that mimic interactions with
a (usually) local neighborhood. The choice of neighborhoods and agent rules can
vary widely and is critical for the performance of the model. While some models,
such as the aforementioned Schelling model, create rules from the intuition of the researcher, models intended for quantitative predictions need rules carefully tuned to data.
This, and also to avoid any bias from ``handcrafting'' rules, requires the 
construction of agents in a more data driven way (e.g. using 
regression approaches, ``population generation methods'' \citep{sun2015bayesian}
or deep learning).  

Compared to human behavior (which includes adaptivity, memory,
sensorimotor skills, social interaction, creativity and more), even
the classical agent models still appear strongly simplified.
However, such simplification
can also be designed systematically in order to investigate which 
behavioral ``components'' (e.g., automatic obstacle avoidance when moving) contribute 
to which aspects of city dynamics (e.g., efficient pedestrian stream dynamics). 

\paragraph{Digital Twins (DTs)} have been developed and successfully used as a
comprehensive planning, analysis, and control concept for complex cyber-physical
systems, e.g., in manufacturing \citep{mihai2022digital}. Subsequently, the same
DT approach was also taken up  for city planning \citep{ferre2022adoption}, another complex area where it is of utmost importance to avoid costly errors and model the (future) reality as closely as possible, despite the numerous interrelationships between the many different aspects of a city \citep{herzogjarkeDTs}. In addition, these model DTs must be capable to model societal necessities such as an emphasis on sustainability and reaching relevant UN Sustainability Development Goals (SDGs)
and to be updatable with not only new data but also new functionalities \citep{herzogjarkeDTs}.

While Digital Twins offer a big step towards agents with rich behavior based on a sophisticated model, there are still major gaps when it comes to a comparison with real humans populating an urban environment, and LLMs offer a potential solution to help bridge such gaps.
\citet{Wang2024} surveys the literature on integrating LLMs and agent technology; several of these approaches (e.g., BabyAGI, CAMEL, Multi-Agent Debate, MetaGPT) are based on multi-agent systems that support collaboration between agents to accomplish a common task. AutoGen \citep{Wu2023} is an open source framework for building multi-agent systems, where agents can integrate humans, tools, and LLMs, or combinations thereof. Agents can use many of the capabilities of advanced LLMs, such as role-playing, state inference, making progress based on conversation history, providing feedback, adapting from feedback, and coding.

The combination of LLMs and agent technology opens up a wide range of promising applications. For example, LLMs can be used to decompose potentially complex tasks into simpler subtasks organized as a workflow. To execute the subtasks, the LLM generates code that is passed to agents, which have the ability to execute tools through code execution  \citep{Li2023,Mans2024}. In addition, agents and LLMs can work together to solve potentially complex optimization or simulation problems, where the LLM generates solver code based on a natural language query and translates the solver result into a natural language response  \citep{Li2023a}. Finally, an agent can assist urban planners in performing complex tasks such as text review, auditing, or evaluation, and assist them in tasks such as reverse geocoding, knowledge graph construction, and image captioning \citep{zhu2024plangpt}.

\section{What AI generative models can do for city models}

Advances in AI and machine learning have greatly impacted on our modeling tools in general. Before these
developments, our only way to create models relied on careful manual design. This usually required significant insights
into the to-be-modeled processes, ideally distilled into mathematical equations to provide the basis for
a computational model. The role of data was limited to identify suitable settings
of the (usually few) parameters of the model to make it fit given data and to predict new data about the phenomenon
of interest. The flexibility of these parameterized models was severely limited by the relatively low number of their
parameters. Increasing this flexibility by introducing more parameters usually led to overfitting and poor
performance on new data. On the other hand, for processes whose inner structure or basic laws were basically unknown,
models with such limited flexibility were very difficult or even impossible to adapt to fit the empirical data. As it
is well known, 
this changed around a decade ago when it was discovered (in the context of models for object recognition from
images \citep{krizhevsky2012imagenet}\footnote{A milestone paper with more than 130000 citations}) that deep neural networks did not pose this overfitting problem of
traditional parameterized models, opening the avenue to what is now known as \textit{data-driven modeling}. 
In this approach, the model emerges almost completely\footnote{The only handcrafted specifications being the choices of the neural network architecture and the training data set plus a handful of hyperparameters.} as the result of training a deep neural network
on a training data set, where usually both the deep neural network and the training data set must be extremely large to make the method work. 

This revolution has greatly pushed forward the frontiers of modeling in many fields, and notably in urban science among them. Here (as well as in many other fields), a particular impact comes from the important subclass of \textit{generative AI models} that can generate complex data objects, such as images,
texts, or also almost any other kind of complex data item, provided that training data sets can provide enough examples \tcg {to cover all relevant knowledge aspects for the urban plan to be created. Therefore, this modeling approach facilitates the overall knowledge acquisition process including also the important interrelations between many urban planning detail tasks. Therefore, this modeling approach, e.g., using specialized Small Language models for urban planning tasks together with Digital Twins, provides a new tool for urban planners that will serve as an "Urban Planning Copilot", since the planning process as such is not changed in its structure. However, generative language models can greatly reduce the human effort for simulation, increase predictive accuracy, enable the exploration of planning variants, provide novel multi-modal interfaces, and  widen and enhance the planning process participation of stakeholders and citizens, also through predictive feedback.   }  

\subsection{Generative Models for images, 3D structures and maps}
Many aspects of urban modeling are related to visualization, a process governed by well-known physics. However, details of surface properties that impact appearance remain cumbersome to model. Here, deep learning models have been able to substitute physics-based modeling with purely data-driven visualization models that learn from large image databases. Most of these models originate from Generative Adversarial Networks (GANs, \citep{goodfellow2014generative}).

GANs provide a very versatile class of data-driven generative models for images
(and other high-dimensional objects) that are generated from having two networks
that learn in an opposing (``adversarial") way from given data (such as an image data base):
a first network (the final generator) transforms its input (in the simplest case some
random vector) into a data sample (e.g., an image). Its learning task is to respond to
each input with a sample that resembles the samples (e.g., images) in the data base.
A second network, the discriminator, receives samples randomly chosen from responses
of the generator or from the database and has to learn the binary decision 
whether they stem from the former or from the latter. Failing to make the right
decision is used as a reward for the generator, thereby driving its learning towards
the production of samples that become more and more indistinguishable from true
database elements.

This technique has now become an almost standard way to enable the generation of complex data items (of which images
are just one major case) from many domains, without requiring any explicit specification of the
desired result except for providing a large database of examples \citep{hong2019generative,gui2020review}. By conditioning the learning of
the two subnets on a shared steering input, the output can be controlled in a flexible manner.
This allows transformations between vastly differing representations of data items to be implemented.
For example, the control input can be image sketches or stylized facade schemas \citep{isola2017image,bachl2020cityganlearningarchitecturalstyles},
which the generator then transforms into photorealistic images. Other recent works demonstrate the generation
of 2D designs of house floor plans (``House-GAN", \citep{nauata2020house}) or
entire 3D structures for buildings (``Building GAN", \citep{chang2021building}) from simple text-based specifications. These techniques have been elaborated in \citep{quan2022urban}, allowing even non-experts to generate entire city street layouts that mimic the urban form of a provided example.

\begin{figure}[ht]
\centering
\includegraphics[width=0.7\columnwidth]{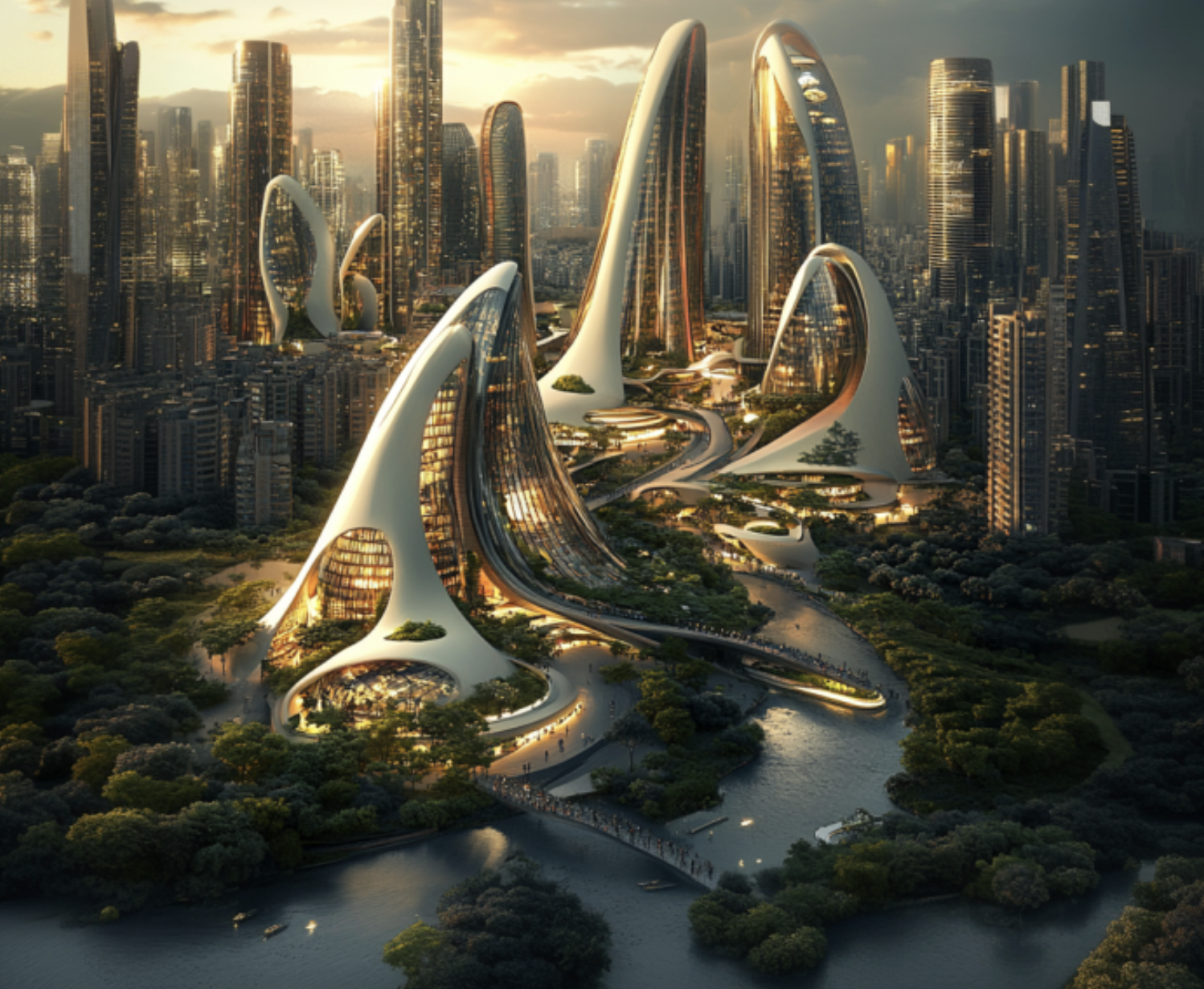}
\caption[caption for LOF]{\small Futuristic city rendered with Midjourney using a specific text prompt\footnotemark.}

\label{fig:futurecity}.
\end{figure}

Recently, another generative approach,  \textit{diffusion models},  has emerged that can produce similar results. It is based on a different principle, exploiting the computational inversion of a diffusion process. The
idea is to generate training sequences, each of which is a sequence of progressively noisy
versions of some different starting image, with each sequence ending when the final image can no longer be distinguished from pure noise.
The generator is obtained by training a predictor network on the reversed sequences. This makes the trained predictor able to iteratively transform even pure noise into an image from the image space used to construct the training
sequences \citep{song2020denoising}. Like GANs, this
approach has become extended with conditioning inputs and further refinements and has begun
to replace GAN generators due to its potential for even higher image quality \citep{dhariwal2021diffusion} --  for a survey see \citet{yang2023diffusion}. Again, it has become combined with front-ends for flexible text-based control of
image generation, thereby often drawing on large language models (LLMs) as pioneered by
GPT and its successors (see below). These approaches are already available in a multitude of commercial (e.g.,  \citet{midjourney}) or free (e.g., Stable Diffusion) platforms
with convenient text-based interfaces to make the image generation process also feasible for non-experts (see Fig. \ref{fig:futurecity} for an example). It is important to note that these results come without any guarantees about physical feasibility since the models were only trained with visual images, not with any deeper representations to capture static or other physical laws.

Besides visualization, data-driven approaches do advance into numerous
other tasks that are important for urban modeling, including generation of land use patterns from
surrounding spatial context to aid urban planning \citep{wang2023automated}, building layout
generation \citep{jiang2023building}, wind flow modeling \citep{kastner2023gan}, floor plan design
from human activity patterns in buildings \citep{wang2021actfloor}, lane extraction from aerial images 
\citep{ruiz2024lanesegmentationrefinementdiffusion}, or generative design of walkable cool areas based on GAN-generated heat patterns \citep{li2024generative}.
Other works demonstrate the design of complex "citiscapes" populated with 3D building models, 
streets and lakes from aerial images \citep{zhang2020automatic}, or the ``CityDreamer" \citep{xie2024citydreamer} that can generate spatially
unbound 3D cities including the rendering of visual appearances for their 3D-buildings.

\footnotetext{``Imagine you are an urban designer. 
Create a view of a city with a harmonic balance
between architecture and green zones. 
The buildings should use curved surfaces
and reflect a sharing of style without being monotonous. Use building smooth,
curved forms that blend well with forms of nature. Make the buildings stand out
aesthetically by giving them bright colors. Render the cityscape in mild late
afternoon sunlight. Avoid skyscrapers and plain rectangularity. Pay careful
attention to each of the requirements."}

\subsection{Large Language Models}

Large Language Models (LLMs) \citep{devlin-etal-2019-bert,brown2020language} are a form of so called \emph{generative AI}. They are neural networks with billions of parameters (the weights of the interconnections between nodes in the network). These parameters are set by training the model of very large amounts of text to predict what might come next given any prompt.  Although simple next word prediction seems a very primitive facility, in fact LLMs have displayed remarkable fluency and ability to answer seemly any question on any topic. They are sometimes termed ``foundation models'' since they can potentially be used as the foundation for an AI system in virtually any domain or application area, perhaps just requiring further ``fine tuning'' or careful prompting strategy (i.e., the precise wording and form of question given to the LLM).  
The performance of LLMs has improved dramatically in recent years with many claims made about their abilities including their ability to reason \citep{creswell2022faithful,huang2023reasoning,kojima2022large}, for commonsense \citep{Belussi-22}, and their ability to pass tests designed to test human ability, 
many point to, possibly fundamental, limitations in their abilities, and suggest that they alone will not lead to Artificial General Intelligence (AGI) \citep{baum2017survey}, but that some kind of neuro-symbolic AI \citep{garcez2023neurosymbolic} might be a more likely  route to AGI; it has already been shown that, for example using LLMs to translate a problem into a logical formulation and then using an external reasoner can improve performance (e.g. \citep{li2024advancing}). The problem addressed in \citep{li2024advancing} was a spatial one based on the StepGame benchmark dataset. The native LLM performance was less good than when the external reasoner was employed in conjunction with the LLM. There have been other demonstrations of the difficulties that LLMs have with spatial reasoning, including the difficulty of reasoning about cardinal directions, e.g., \citep{cohn2024evaluatingabilitylargelanguage}, particularly 
 intercardinal directions (NE, SW, NW, SE).

\begin{figure}
\centering
\begin{minipage}{.485\textwidth}
  \centering
  \vspace*{-0.7cm}
  \includegraphics[width=.99\linewidth]{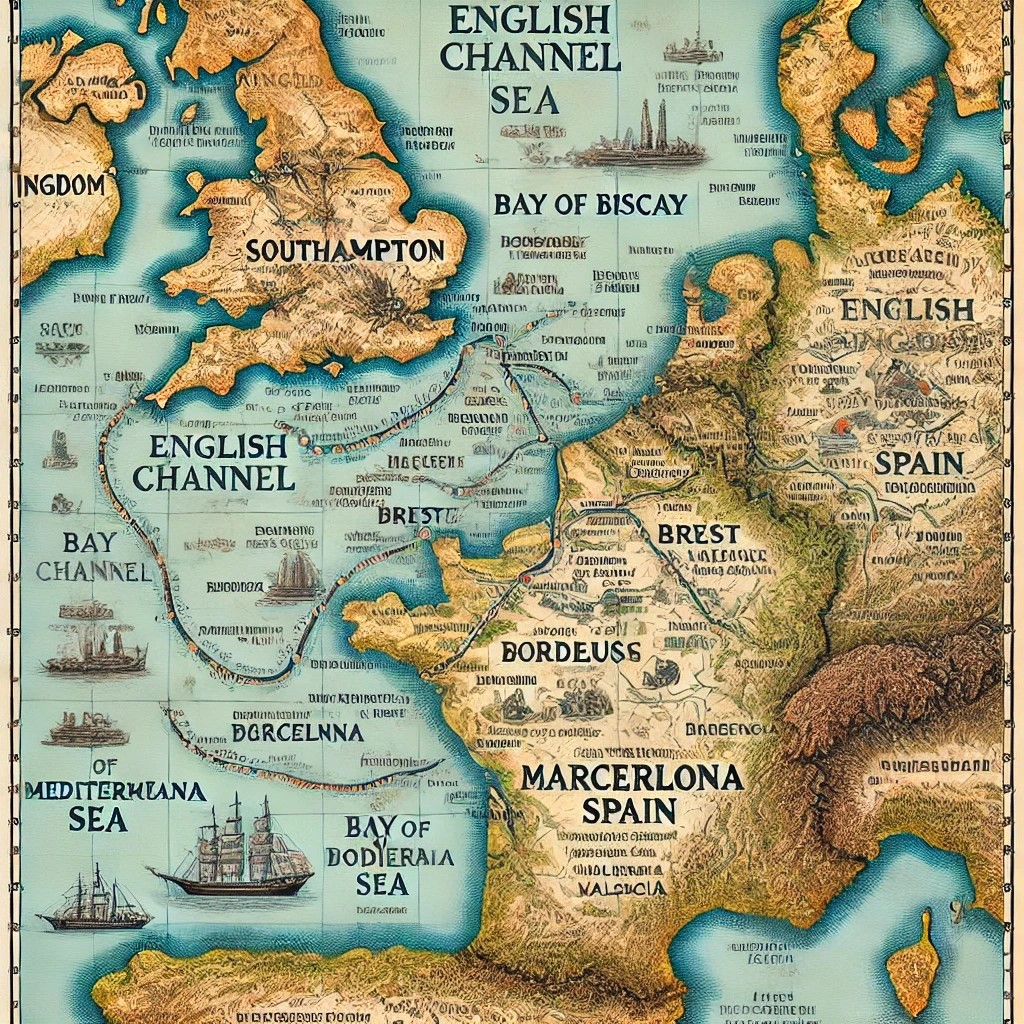}
  \captionof{figure}{The image drawn on 3/9/24 by ChatGPT-4o to the prompt ``give me an image of a map of the sea route from Southampton to Barcelona.''}
  \label{fig:southampton-barcelona}
\end{minipage}%
\begin{minipage}{.03\textwidth}
  \centering
  \ 
\end{minipage}%
\begin{minipage}{.485\textwidth}
  \centering
  \includegraphics[width=.99\linewidth]{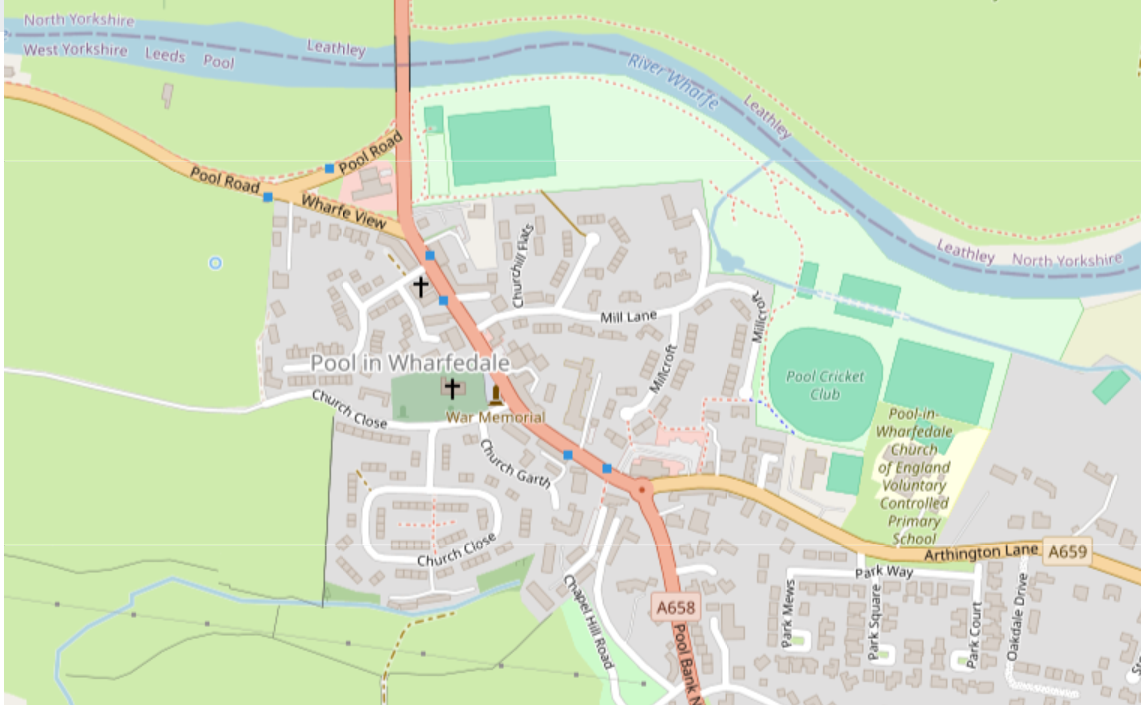}
  \captionof{figure}{A map of Pool-in-Wharfedale in West Yorkshire, UK. Presented to ChatGPT-4o as a raster image taken from Open Street Map.  The given prompt on 4 Sept 2024 was: ``Consider the attached map.  Please describe the main geographical  and built environment features''.  }
  \label{fig:pool-in-wharfedale}
\end{minipage}%
\begin{minipage}{.03\textwidth}
  \centering
  \%
\end{minipage}
\end{figure}

A more recent development is the emergence of multi-modal LLMs (MLLMs) -- which can for example, process and generate images as well as just text\citep{zhang2024mmllmsrecentadvancesmultimodal}.  
A recent survey is \citet{zhang2024mmllmsrecentadvancesmultimodal}.  In principle, multi-modal LLMs include other kinds of modalities, for example speech input and generation, but we will just consider vision-language MLLMs here. There are two aspects to consider: (1) MLLMs which can generate images given text prompt and (2) MLLMs which can accept images or other kinds of visual inputs and analyse them, and then respond to textual prompts.  
We have already seen the impressive abilities of systems such as Midjourney  and DALL-E to generate images to order in Fig. \ref{fig:futurecity}.  
However, their map drawing abilities are very poor -- see, for example, Fig. \ref{fig:southampton-barcelona} -- there are so many things wrong with the map including the fact that the requested route is not shown that one does not know where to start.

MLLM's capabilities regarding the analysis of images and maps in particular have been improving. For example, consider the  map displayed in figures \ref{fig:pool-in-wharfedale} 
and the descriptions generated by ChatGPT-4o which can be found in the appendix as Fig. \ref{fig:pool-in-wharfedale-output}.
The description is generally very good and accurate, though it contains a few subjective opinions or surmises about the region not backed up by the image (e.g., the School being an important community institution, and that the bridge is a crucial piece of infrastructure) -- both are in fact true but cannot be inferred directly from the map -- though it is possible the LLM used information gathered during training to make these statements.  

However, when it comes to interpreting sketch maps LLMs seem to struggle more  -- see Fig. \ref{fig:sketch-map} and the description given of it in the Appendix, Fig. \ref{sketch-map-output}. Some parts of the description are good, in particular all the text relating to the roads -- except for the purple road which is ignored in much of the output.  However, the locations of the buildings are mostly wrong.  Moreover, if asked to plot a route from the Fishmonger to the School, many mistakes are made (e.g., one should not go past the newsagent, one should turn right not left on to the red road -- see Fig. \ref{fig:fishmonger-school-route}).

\tcg{These examples exemplify very well that the problems of the verification or even only validation of LLMs is still an open problem which is not astonishing: for "traditional" program systems, exhaustive tests or even their verification is still state-of-the-art only for few high-profile systems. Given the ambiguity of natural language prompts of LLMs and the possibly ambiguous interpretation of multi-modal input, the difficulties for developing verification or validation methods are exacerbated compared to "ordinary" program code. In order to build upon a valid base for an "Urban Planning Copilot", the quality of the training or finetuning data is of utmost importance: the data must be clean, diverse, and must cover all urban planning scenarios. The next step would be to test a trained LLM under various conditions, including adversarial attacks, to ensure its robustness and reliability. However, during its use, it is necessary to continuously and interactively monitor an LLM model's performance while developing urban plans. This can lead to performance metrics through an LLM model evaluation of its accuracy, precision, recall and F1 score which are performance measures taken from natural language processing, e.g., from full-text search.} 

\tcg{This validation and verification problem is another argument against LLMs and for Small Language Models: a highly specialized Small Language Model can be much better dealt with compared with the complexity of an all-encompassing LLM.}

\begin{figure}
\centering
\includegraphics[width=0.7\columnwidth]{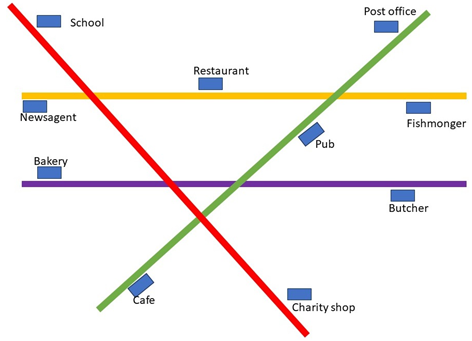}
\caption{A sketch map given to  ChatGPT4-o on  3rd September 2024  with this instruction : ``Please consider the attached sketch map and describe the various features on it in sufficient detail that someone else could create a reasonable copy of it.'' See Fig. \ref{sketch-map-output} in the Appendix for the LLM's response. }.
\label{fig:sketch-map}.
\end{figure}

\subsection{Geo-LLMs}

In the previous section we looked at LLMs generally and their abilities to reason with or generate spatial information and maps.  In this subsection we report on some of the growing body of research into geo-LLMs.  For example, GEOLM \citep{liGeoLMEmpoweringLanguage2023} aims to connect geo-entity mentions in text to  geospatial information extracted from geographical databases; the authors demonstrate that GEOLM has some capability to  support toponym recognition, toponym linking, relation extraction, and geo-entity typing.
\citet{Qi2023} propose that rather than using LLMs \emph{for} spatial databases, they should be used \emph{as} spatial databases -- ``Machine Learning Models as a Spatial Database'' (MaaSDB).  Whilst they point out the advantages this might bring (learning from and integrating information on  both structured and unstructured data; full integration between the natural language interface and the data itself), this is not without risks -- LLMs are well known for their propensity to ``hallucinate'' \citep{huang2023survey} and thus relying on an LLM for accurate spatial fact retrieval may yield incorrect results (and indeed the paper itself gives examples of incorrect information being returned in a small number of cases in their preliminary study using ChatGPT).  Other work investigating geo-spatial capabilities of LLMs includes \citet{bhandari2023largelanguagemodelsgeospatially} who conclude that at least the largest LLMs, and those that have been instruction-tuned do have fair abilities. 

An interesting approach to interacting with geographic data extracted from a map is presented by \citet{unlu2023chatmap}. In this work, a number of points were chosen such that  300 metre radius circles centred on these points were densely tagged with features (e.g. schools, cafes, ATMs...).  The amenities  of each area are then represented as a \emph{preprompt}. E.g.:
\begin{quote}
   \emph{This is a circular area of radius of 300 meters that intersects province(s) of İstanbul
and district(s) of Fatih. There are 3 atm(s), 2 bank(s), 1 bureau\_de\_change(s),
18 cafe(s), 2 clinic(s), 1 court\_house(s), 2 dentist(s), 1 driving\_school(s), 2
events\_venue(s), 11 fast\_food(s), 1 guest\_house(s), 3 hospital(s), 11 parking(s), 33
pharmacy(s), 9 place\_of\_worship(s), 1 post\_office(s), 43 restaurant(s), 5 school(s),
1 shower(s). There are 525 buildings which cover 31\% of the total area. It contains
289 meters of platform rail, 100 meters of footway road, 80 meters of pedestrian road,
44 meters of primary\_link road, 2786 meters of residential road, 283 meters of service
road, 20 meters of steps road, 1005 meters of tertiary road, 62 meters of tertiary\_link
road, 249 meters of unclassified road.}
\end{quote}

ChatGPT 3.5Turbo was then used to generate prompt answer pairs from these preprompts, using prompts such as:
\begin{quote}
    \emph{I will give these types of preprompts and you will generate prompt-answer pairs in
python list of dictionaries format. These prompts should be questions that businessmen,
citizens, tourists would demand based on the data in the preprompt. Generate 50
prompt-answer pairs with very diverse topics. Important : Do not generate prompts
that data in preprompt is not sufficient to answer !}
\end{quote}
ChatGPT 3.5 Turbo then produced prompt-answer pairs such as:
\begin{quote}
    Question : Tell me about the options for cultural enthusiasts with a gallery nearby.\\
Answer : Cultural enthusiasts can explore the gallery in this area.
\end{quote}

4111 such prompt-answer pairs (generated from 81 preprompts) were then used to fine-tune an LLM  (Falcon 1B RW) using Low Rank Adaption (LORA) \citep{hu2021lora}. The resulting fine-tuned model is able to answer questions similar to those in the fine-tuning set about novel regions.

\begin{figure}[ht]
\centering
\includegraphics[width=0.99\columnwidth]{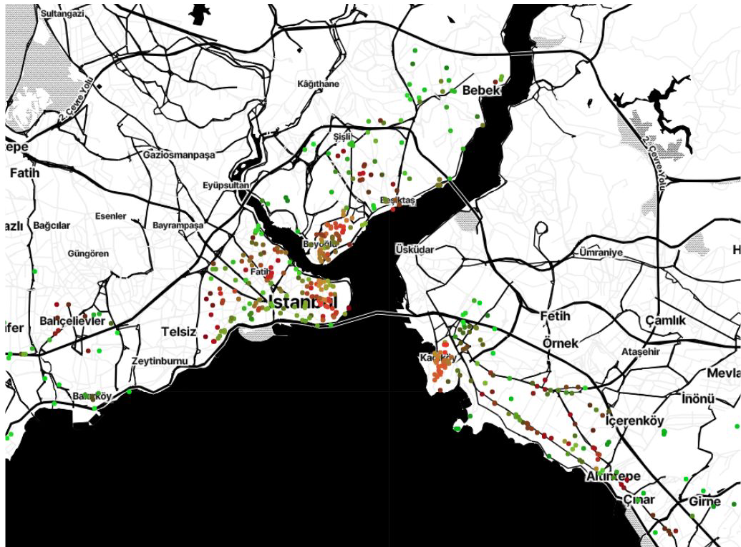}
\caption{Embeddings of preprompts visualised on a map where colours are assigned proportionally to UMAP values. ``Bright red colors indicate more touristic locations, dark red colors
more indicate business/commercial districts and bright greenish colors indicate relatively empty spaces, residential areas'' \citep[][Fig.4]{unlu2023chatmap}.
}
\label{fig:chatmap-map}.
\end{figure}

A further experiment reported in the paper is to map the (GLOVE) embeddings of the preprompts onto a ``D space using universal manifold approximation and projection (UMAP) -- and then assign colours to each embedding based on its UMAP score -- these can then be plotted on a map as shown in Fig. \ref{fig:chatmap-map}.

\citet{mai2023opportunities} discuss some possibilities for exploiting Foundation Models in geospatial situations, investigating the performance of a number of different models across seven tasks.  They conclude that on toponym recognition and location description recognition (which are purely textual tasks), LLMs can outperform task specific fully supervised models. On the other hand for multi-modal problems, such as land classification from satellite imagery, or street view image-based urban noise intensity classification, specialist models outperformed multi-modal foundation models. They conclude that multi-modal models specifically built or tuned for geospatial data should be an aim for future work.  

SpaBERT \citep{li2022spabert} is a spatial language model based on the BERT  LLM \citep{devlin-etal-2019-bert}. Just as standard LLMs use embeddings based on neighbouring words in sentences (which SpaBERT continues to use), SpaBERT also builds embeddings based on neighbouring  entities in geographic space; these are linearised based on the distance to the \emph{pivot} entity; SpaBERT also builds a third embedding based on the distances to the nearest neighbours in the $x$ and $y$ dimensions separately.  SpaBERT is then pretrained  in a standard way with masked language modeling and masked entity prediction tasks to learn spatial dependencies. When applied to two downstream tasks: geo-entity typing and geo-entity linking, SpaBERT shows significant performance improvement compared to standard LLMs without any specific spatial competency.

\subsection{LLMs and Retrieval Augmented Generation (RAG)}
We have already mentioned above the problem of ``hallucination'' -- that LLMs tend to produce text which contains statements which are not true -- see \citet{hongHallucinationsLeaderboardOpen2024} for an open effort to measure the extent to which this happens via a series of benchmarks.  There are a number of ways  in which this issue might be mitigated, although it is unlikely that hallucinations can be eliminated  - indeed \citet{banerjeeLLMsWillAlways2024} argue that they are something we need to live with, arguing that  ``every stage of the LLM process -- from training data compilation to fact retrieval, intent classification, and text generation -- will have a non-zero probability of producing hallucinations''. Similarly \citet{xuHallucinationInevitableInnate2024} 
use results from learning theory to demonstrate that LLMs will  always hallucinate.
However, there is a community focus on finding ways to mitigate the problem, recently surveyed by \citet{tonmoyComprehensiveSurveyHallucination2024}. They outline a taxonomy of possible methods, with the two top-level categories being prompt engineering and model development.  Within the former category, Prompt Tuning, Self-Refinement and RAG form the next level of the classification.  RAG is of particular interest in our domain of interest. RAG addresses one of LLM's biggest weak spots -- since LLMs are only (re)-trained rather infrequently -- because of the huge cost -- so their ``knowledge" may date quickly, particularly about propositions  whose truth  may change frequently. RAG provides a way for LLMs to tap into authoritative symbolic knowledge bases as a secondary explicit knowledge source rather than relying on information held implicitly in its model from possibly outdated training data. This symbolic knowledge can be exploited in three different ways by a RAG enabled system: (1) the external knowledge can be added to the user-supplied prompt before generation starts; (2) during generation, for example by validating generated facts and rectifying them when necessary; (3) waiting until an entire response has been generated before checking it against external knowledge to the LLM  and adapting the responses as necessary.

\section{The future of urban models: prospects and opportunities}

Cities are for citizens. The introduction of agent-based modeling techniques has allowed  for significant strides towards more realistic models, but the created agent models were still severely limited by the comparably simple behaviors that could be realized in those agents.

\subsection{{What is missing?}}

So far, urban planning has been mainly concerned with the static infrastructure of cities. Adding
virtual
citizens and allowing them to move and behave
in their environment according to their needs is a very important extension. 
This can  be built using  the agent-based
modeling techniques described above, but it is an open question whether these modeling methods can lead to agents
that have 
sufficient complexity to yield systems with an adequate level of realism. 
The simple parameterised agent models that aim to be ``statistically correct'' with regard to their aggregated properties (such as means and correlation functions, see, e.g., \citep{sun2015bayesian}) can indeed capture some useful aspects
of behavior to drive models such as transport systems under loads of simulated passengers. However, these models usually attempt an approximation at a certain time scale, with the agents deriving their behavior from relatively simple sensor signals and following mostly fixed parameterised rules to compute their movement decisions or trajectories. While well calibrated models of that type have shown useful predictive capabilities for real systems, they are limited in many ways, e.g., they cannot deal well with changing conditions, such as unusual weather
conditions, emergency situations or simply drifts due to technology changes
(e.g., increasing numbers of cars, with changing driving characteristics due
to partly or fully autonomous driving systems, changes in car utilization
due to sharing or home office). If at all, they have only very
simple goals, intentions, emotions, adaptivity or creative capabilities. This exposes a stark difference between current agent models and real citizens.

\paragraph{\textbf{Citizens have mind states.}} 
The behavior of citizens can be extremely rich, reflecting their largely unobservable  states of mind that make them extremely adaptive and able to correlate their behavior with memories from decades ago. They enable citizens to shape and take highly informed initiatives, to be emotional, have expectations and values and to be creative. As a result, they can engage in rich interactions among themselves, making reciprocal representations of each other, and interact
similarly with the
city in which they live. From the observable, outer side, this fills
the city with complex interaction patterns that actually underlie most
of its development. In addition, the city becomes connected and embedded in the mental spaces of its citizens. In turn, these mental spaces become filled with experiences that arise from the interactions with others and with what the city has to offer to them. This shapes needs, demands, and resources that reshape mind states, changing the complex relational fabric that the citizens weave into the city. It extends the city's physical space with a rich social space in which the citizens' mental spaces are connected and interrelated. While this social space (along with the individual
citizens's mental spaces) can only be observed indirectly, by inferring some of their content  from their shaping of what the citizens do (e.g., usage of media), it is an entity that is at the root of almost all impacts that the citizens bring to bear on the city.
\medskip

In the following, we take a brief tour through several of these
aspects.

\paragraph{\textbf{Capturing rich knowledge.}}  Human behaviors obtain their richness from the huge amount
of knowledge that affects human choices. This knowledge has a complex 
and hierarchical structure: the largest part -- our semantic memory -- has been
distilled from lifelong learning from huge numbers of contexts. Another part
is from our episodic memory linked to specific episodes in our past. And a
third part of our short term and our working memory captures very recent
contexts and events just minutes, hours and days ago. 
Capturing such rich and 
hierarchical knowledge requires new methods that have opened up powerful ways of creating models completely from data, leading to new types of generative models culminating in the LLMs epitomized by GPT and its successors. The next section \label{ai-impact} is dedicated to providing a compact review of these new
possibilities. Finally,  we \label{future} will attempt to look at how this opens up exciting perspectives to complement modern agent models such as Digital Twins with powerful capabilities to bring city models to a new level.  

\paragraph{Integration of demands and resources as key drivers.}
Citizens are both the carriers of resources as well as of demands.
A city is attractive when it supports a good matching between resources
and citizens' demands. Therefore, we need models that allow 
for
the introduction of
resource-demand matchings 
into the planning process.
Such matchings have already been considered as important
elements of transportation models \citep{zhang2021agent}. Matching processes that
are focused on local neighborhoods are -- in highly idealized form --
also at the core of the cellular Schelling model \citep{jensen2022introducing}, in which residents
match their expectations about the ethnic composition of their
neighborhood with possible choices for a picking their residence.
This matching model has been shown to explain the strong tendency
of ethnic group mixtures toward spatial segregation. 

The importance of linking the needs of the citizens with the urban planning processes has been hardly discussed in the literature \citep{alsayed2024urban}.\citet{gonzalez2021agent} provides
one recent example, aiming towards an agent-driven platform that can
mediate between different stakeholders in the process. 

\paragraph{\textbf{Smart support environments.}} The now wide-spread availability of human-compatible perception and natural language capabilities in technical artefacts and the ability to
compile huge amounts of human expertise into digital assistants
leads to the question how to combine such skills with human
expert guidance. This requires to modify standard deep-learning
architectures to provide for meaningful inclusion of human expert guidance, e.g., when realising intelligent urban planning assistance functions \cite{fang2022incorporating}. It is also a question to the field of human-computer interaction, extending its scope from 
artefacts surrounded by humans to entire environments inhabited by humans. For the case of buildings this has spurred human-building-interaction (HBI) as a new branch of HCI \cite{alavi2019introduction}. As the next logical step we foresee its extension to the even more comprehensive field of human-city-interaction, with the task to elucidate the integration
of digital and AI-driven technology with the planning, maintenance and operation of urban structures and services in ways that enhance urban quality and adress human needs. 

\paragraph{\textbf{Citizens' mental spaces.}} Human interactions are not only highly variable, they also add a level that is
not present in physical interactions that are familiar from the physical
city infrastructure. This level is constituted of \textit{mental spaces}
that are inseparable from human agents. Humans not only feel  anchored 
in the physical space of the city. They also augment it with their mental space
which gives them imagination and the power of abstraction. They can populate their
mental space with their individual versions of the city or parts thereof, e.g., 
creating their individual version of the subset of city places which they
know and their preferred commuting trails between them \citep{gibala2016real}. They can create mental
pictures of ``their ideal city" and their life therein, or mental maps of safety
zones, recreation areas, landmarks or networks with other citizens. 
And they can open further spaces accommodating
their needs, hobbies, dreams and desires. All these
possibilities are ultimately rooted in their experiences in the physical city,
but extend its physical space and feed back on it in rich and varied ways \citep{mehaffy2023smart} that reflect their rich
relationships with the city they live in. Can we build models that represent
aspects of these mental spaces that are shared among large groups of citizens?
How would these models interact with the physical city? How do the physical
and the shared mental city spaces jointly affect decisions, identity or creativity
of the citizens? All these are modeling challenges for the future almost untouched
so far. 

\paragraph{Social space.} Of these mental spaces there is a particularly 
important one that is specialized on the social dimension of human 
interactions and, therefore, deserves special attention.
The social nature of human interactions becomes already relevant when a human
interacts with an artificial agent, such as an avatar in a mobile phone or 
an embodied robot. The human will locate the agent not only in physical space,
but also in social space, perceiving the agent, e.g., as representing an external
agency or alternatively as a member of the human's household. This leads to different
expectations about agent role, trustability, party taking in decisions, or
compatibility of interests and intentions. This requires a ``\textbf{Social AI}'' that
extends beyond the merely functional level and that can
take social aspects into account in its interactions and services.
Conceptually, this again
can be framed as the augmentation of the city with another space that represents
the social interrelations among the citizens.
Here too, we may distinguish between
a ``shared approximation" of what might be termed ``objective social space",
and individual spaces (as specialized versions of the individual social spaces of
citizens) that represent the individual perspectives of citizens' own social
embedding in the city.

\paragraph{\textbf{Social AI}.} While traditional city models were focused primarily on the functionality
of the infrastructure of a city, with subsequently developed agent models 
adding to that level a layer of behavior in the form of (usually highly
simplified) decision-making (including decisions about agent movements, e.g.,
when the agents model pedestrians or drivers), we are now at the dawn of models
that can include rich social behavior, identity and creativity 
enabled through the presence of these social and further mental spaces.
For instance, geotagged twitter data have been used to create models about demographic variations of distancing practices during the
COVID-19 pandemic \cite{xu2024twitter}. Commuting efficiency \cite{ling2024unravelling} quantifies how well urban policies optimize the
connection between job and residential locations, and how the degree of optimization varies depending on socio-demographic factors.
The path towards smart cities leads to an increasing multitude of platforms to capture data about such and numerous further aspects of citizen life \cite{pal2018big}, providing a rapidly evolving basis for
more detailed models of citizen behavior.

This will connect city
models with the concept of a ``Social AI'' as a key element to adequately frame
the rich interactions of citizens as social entities in interaction with each
other and their mental spaces, and also with AI agents that will become an 
increasingly pervasive part of
the future infrastructure of cities, making cities accommodate a hybrid society
of humans and artificial agents which can interact in new synergies to drive the
emergence of our cities of the future. 

\paragraph{\textbf{Interdisciplinarity.}}
The above makes clear that a proper addressing of the complexity of urban models necessarily
has to cut across many disciplines:

Since cities connect the lives of their citizens, appropriate models
have to cope with tasks that involve numerous areas in addition
to urban and computer science. To model the forces and dynamics
that unfold when citizens meet and move in cities needs to reach out
into fields that deal with us humans: psychology, sociology, 
health sciences, arts and communication, and what it takes to organise
us: education, economy, law, and certainly many more.
Neither last nor least in this enumeration is ethics: modern cities
and their data- and AI-driven models have many of their benefits 
connected with significant ethical questions concerning privacy, freedom,
forms of governance, ownership, values, or norms.

\paragraph{\textbf{Case studies.}} Resources and methods for data-driven
models and modern AI methods working at the scale of cities 
are rather recent developments. Thus case studies typically were
initially restricted to major metropoles and will certainly
become more numerous within the next years.  
The case study \cite{Sanchez_Sepulveda_2024} for the city
of Barcelona provides an early example of the use of data-driven models
to inform city planning with a focus on improving walkability, transportation
and safety. Another example \cite{floridi2020artificial} presents insights and benefits 
for the cities of Amsterdam and Helsinki when
making AI services accessible through centralized AI registers. 
The case of the city of Singapore \citep{das2024ai} illustrates
how addressing the needs of the COVID pandemic benefited from AI-driven
approaches to realize different subsystems that became gradually more and more
integrated and were adapted to serve new tasks after the pandemic, leading
to a seamless merging of IoT and AI to explore different ways of ultimately making
a city adaptive in real-time, e.g., through smart gantries or police robots
perceiving and reacting to human behavior to enforce policy rules.
Typically, case studies can only address a subset of the wide range
of AI use cases that are relevant to a city. Focusing on the transportation sector, \cite{patil2024artificial} provides an
overview with a taxonomy of major case studies obtained from an extensive literature 
search on AI applications in urban contexts.
A very large use case has been launched around 2017 in the Haidian district
of Beijing \citep{cugurullo2024ais}. After a concept phase and the formation of  a consortium
of firms and governmental institutions the implementation was started in 2020.
So far, the project has connected more than 14,000 cameras with a large
network of server nodes and two centralized data centers that are running
algorithms and AI methods to process the data to provide a layer of 
automated city management based on combining AI methods for visual
recognition and event prediction akin to a ``brain" that must run
a body that is in this case a city. This approach has inspired
an increasing number of similar ``city brain" projects for other
Chinese cities. Besides a general feasibility of the approach, it has
led to the insight that different cities may require different
tailored "brain structures" and that adequate participation by the
residents remains a major challenge for the current organisational format.

\subsection{Realizing rich virtual citizen models}

We contend that LLMs put us in a position to create agents that can draw on extensive knowledge for choosing their actions, and in a way that can take complex context into account. Moreover, the nature of LLMs provides these agents with an intrinsic form of short term memory that makes their responses contingent on their interactions within a recent time horizon that be made to include dialog spans of many thousand items if desired. This makes these agents quick to adapt to their recent past, a property which has at best been extremely simple and limited in agents used by previous techniques.

These agents allow city models to be brought to a new level where the city is modeled \textit{together with the interactions with and among its virtual citizens}, which can access the facilities of virtual city models in a knowledgeable and adaptive way. 

This will open up entirely new possibilities for urban planners: they can devise different city layouts, populate them with different mixtures of artificial ``agent personas" (which can be quite easily obtained with combinations of prompting and fine-tuning techniques), and simulate the resulting dynamics to predict the impact on citizens and explore and evaluate different designs.

First steps towards such models already exist. \citet{Wang:2024aa} show the realization of an LLM-agent based on GPT-3.5 that can create personal mobility data that then can be tuned with available human mobility data. They demonstrate that their citizen emulation leads to appropriate activity changes when the agent receives a prompt that informs about the presence of a pandemic and pertinent government advice. \citet{park2023generative}  show that a group of simulated citizens placed in a small neighborhood of a simulated city
area unfold a variety of interesting behaviors in their environment, such as going for a walk, making appointments, meeting at a café, or sharing news with a colleague, demonstrating that LLM agents can model complex social interactions and decision-making processes. Besides
enriching models with realistic citizen behaviors LLM agents can also be employed
to contribute to a variety of city planning tasks directly, for instance, by using fine-tuning of base models with scientific literature to create experts for specialized areas, such as urban renewal \citep{Wang:2023aa}.

These possibilities have triggered a significant number of proposals for integrating LLMs with agent technology \citep{Wang2024}. Some of these proposals are based on single-agent systems, such as AutoGPT or LangChain Agents, while others employ a collection of collaborative agents, such as aBabyAGI, CAMEL, or AutoGen. In the latter case, the agents may incorporate individual LLMs to support collaborative task solving \citep{Wu2023}.

An attractive feature of these LLM 
integrations
in architectures that
allow the flexible interfacing of LLMs with additional functionalities (e.g., sensors for perception, databases for specialized information queries,
or actuators for allowing physical or virtual actions) is to give
the LLM the possibility to \textit{access existing tools} to support operations 
for which a variety of established solutions of proven quality exist. This permits the LLM to focus on solving problems at a higher level of abstraction, away from the details of the more specialized operations covered by the tool. In essence, LLM support can be limited to combining existing functional building blocks into a workflow to solve the given task. 

For instance, as we have seen, current LLMs still have difficulties with
many spatial tasks. Thus, basic spatial functions, such as querying a GIS
database or visualizing spatial data on a map, can be safely delegated to a
tool, such as ArcGIS \citep{cityengine2024}, while the task chain for solving a complex task using these basic functions might be generated by an LLM. 

For example, \citet{Li2022} proposed a single-agent approach for performing GIS tasks. An LLM embedded in an autonomous agent performs the planning task. The agent uses basic GIS functionality such as spatial databases, spatial analysis functions or visualization tools together with existing GIS data to perform its task.  The LLM establishes a work flow, generates code to call basic GIS functions and to assemble the final result.

\subsection{Towards LLM ecosystems}
However, the very high computational and memory demands of the currently best LLMs makes their use within such LLM agent frameworks extremely compute intensive and puts severe limits on the number of agents that can be simulated within reasonable resources. Using these ``huge" LLMs to train smaller LLMs has enabled the ``distillation"  of their knowledge into much smaller footprints, giving rise thus to large number of ``small LLMs'' (SLMs) \citep{zhou2023minigiants}. Despite their much smaller size the most recent of these SLMs often sacrifice only  little of the performance of their much larger brethren. At the same time, they can be tuned (or sometimes even trained from scratch) with much more modest resources. Their execution is usually possible, even with several instances simultaneously, on desktop computers down to laptops or even mobile phones \citep{chen2024octopus}. Besides clever training schemes and the use of carefully curated data, their compact size has also been enabled through more compute efficient architectures and algorithmic improvements (``weight quantization'') that allow them to work with extremely low-resolution parameters replacing standard 32-bit float values for network parameters by coarser number representations that occupy only 3-4 bit, bringing a ten-fold reduction in required memory size. Recently, this has been pushed to the extreme of ternary (-1/0/1) weight and binary (0/1) activity values. With suitably adapted algorithms for training and execution these models only require additions and allow for a very energy parsimonious execution on dedicated hardware (such as standard FPGAs) \citep{chen2024ternaryllm}.

The availability of such very parsimonious LLMs opens up \textit{the vision of creating systems populated by large numbers of SLMs} 
to simulate correspondingly large numbers of citizens for a new generation of \textit{bottom-up driven urban models}, that at the same time \textit{are connected to our high-level semantic world}. 
Such a system could then bring the ``mental citizen spaces" into the urban model and simulate the interaction between
the physics of the buildings and the mental spaces of the citizens.
While such modeling will always be too coarse to obtain predictions about mental spaces of individuals, it may be
useful to address phenomena such as the clustering or spreading of opinions and sentiments, their relation to spatial structure (e.g., well-connected vs. scantly connected areas, or local ethnic variations in the city) or to changes in the urban infrastructure (e.g., new apartment and shopping complexes,
roads, or park areas). 

Alternatively, we can realise intelligent distributed ecosystems that can be \textit{embedded in a city}, with each node providing
expertise for a restricted area. The nodes in such a distributed system could be much more easily maintained since the burden of retraining or fine-tuning becomes partitioned into much smaller chunks.
At the same time, such an approach would offer better resilience, since such a system has at worst only reduced functionality if a sub-unit fails, whilst a monolithic unit can only fail catastrophically. 

This would also facilitate the application of RAG techniques to swap in locally
specific knowledge \citep{lewis2020retrieval,fan2024survey}, and 
\textit{cooperation with Digital Twins (DTs)}: DT instances typically represent
models that are in correspondence to many kinds of physical subsystems in a city
(such as road segments, buildings, or vehicles), which makes them tightly
integrated into the ``physical fabric'' of a city \citep{ferre2022adoption}.

\begin{figure}
\centering
\includegraphics[width=0.6\columnwidth]{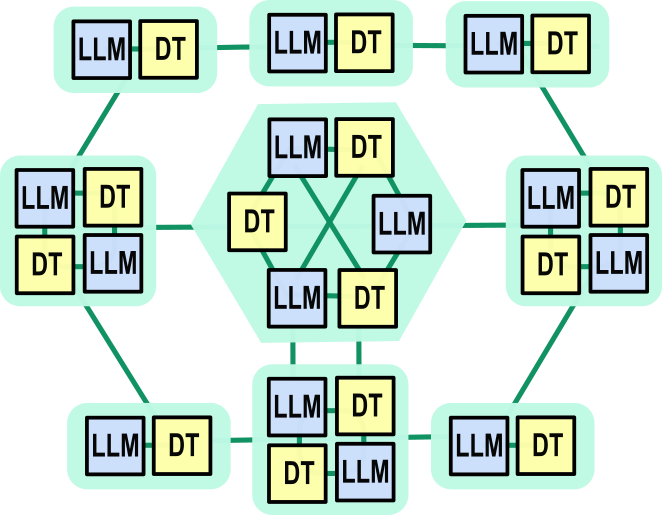}
\caption{Interconnected pairs of locally specialized Digital twins (DT) and language models (LLM) can provide powerful units for a distributed ``intelligent ecosystem" within a city.\label{fig:i_ecosystem}}
\end{figure}

The flexible granularity of SLM arrays offers an ideal match for this granularity of DT systems, allowing each DT to become coupled with one or more  dedicated SLMs from the array. In this coupling, the use of an SLM can extend well beyond just merely providing a natural language interface. Instead, the good tunability (or even trainability from scratch) of SLMs allows the interface to be \textit{knowledgeable and specialized} to the task domain of the DT it is associated with (Fig. \ref{fig:i_ecosystem}).
Such \textit{knowledge-integrated} DT-SLM-pairs offer a new and very powerful building block to endow a city with a kind of virtual and distributed \textit{``intelligent ecosystem"}. Each node of this ecosystem integrates subsymbolic control knowledge in the DT with natural language accessible symbolic and high-level knowledge in the associated SLM. This network can be hierarchical, with super-ordinate nodes supervising more local clusters. However, the selection of collaborating partners in such a large-scale system has been identified as a challenging problem \citep{Rothermel2025}. In the IoT domain, collaboration partners are often selected based on spatial proximity. This is in line with Tobler's observation \citep{Tobler1970} that ``everything is related to everything else, but nearby things are more related than distant things". In \citep{Rothermel2025}, we propose a proximity model that allows agents to identify collaborators based on spatial proximity. However, the concept of proximity is not limited to the spatial dimension and could therefore be extended to cognitive, institutional, social or organizational aspects. 

Combining SLMs and Digital Twins and embedding them in a hierarchical network structured according to the spatial proximity structure of the city leads to an architecture which is very favorable for computational parsimony. While a more detailed analysis has to be postponed to a forthcoming paper, there exist already ample results how the utilization of
SLMs can lead to significant computational gains. For instance, \cite{irugalbandara2024scaling} find in their study of 29 recent SLMs that
most of them reach response qualities that are close to GPT-4 level while incurring only a small fraction (1/5 or less) of the cost. The recent development of scalable matrix-multiplication free language models brings further reductions into reach. \cite{zhu2024scalable} reports an FPGA implementation of a 1 billion parameter SLM employing this technique that
consumes 13W of power while generating text at humen reading speed.

The implementation of such an SLM-DT based ``intelligent ecosystem" can provide a powerful infrastructure for planners to interact with the city, to monitor city states, or even to simulate the response of the city to
unforeseen impacts. Its accessibility through natural language also enables its use to capture feedback or sentiments from citizens, to engage in distributed discussions about planning options, or to support forms of participatory planning that involve citizens via the interface.
Unlike traditional forms of citizen participation, such a network might be able to evaluate the consistency of citizens' proposals, oversee compatibility with overarching long-term plans (e.g., such as the example of the Shanghai Masterplan \citep{shanghaimasterplan}), or evaluate feasibility with regard to resources and demands of involved players.

\subsection{Ethical Challenges}
Urban planning creates long lasting impacts that affect the lives of many. This brings a large responsibility. At the same time, the multi-dimensional optimization expected from urban planners cannot be done without weighting and balancing a multitude of often conflicting optimization criteria and interests. This can lead into ethical dilemmata whose resolution will be affected by the tools that provide planners (and the public) with the models, simulations, perspectives and predictions for different planning scenarios and potential outcomes. A recent
overview of major challenges is provided by \citep{sanchez2024ethical}.
They focus on the problems of bias, privacy, equity and inclusivity,
accountability and transparency, danger of mis- and dis-information,
and the necessity of human control and oversight to safeguard against
limitations of AI, and they offer several recommendations in each of these
areas. \cite{kitchin2016ethics} takes a critical look at how the transition from small to big data influences our way how we conceive cities and their operational governance, with the danger of a ``reduction of urban life to logic and calculative rules". He argues for a re-orientation in how cities are conceived and, likewise, emphasizes the
importance of an ethics about privacy, harms, notice and consent, and recommends a pro-active stance of city managers. We would like to complement these broad analyses with three specific aspects that appear to us particularly focal to our work:

\paragraph{\textbf{Offering intelligent agents roles of virtual citizens.}} Here, the created benefits come with the side-effect of 
potentially substituting real human action or feedback with that from a digital surrogate
that may introduce (or even enhance) biases, compromise privacy (when the simulated behavior leaks individual information in the training set),
or possibly endanger human participation and trust. This will require
advances on the technology side in conjunction with
clear legal frameworks to ensure that citizen models are based on training data that properly reflect local diversity patterns (e.g. gender, cultural, age) and are ethically sourced.

\paragraph{\textbf{Pervasive urban AI with perfect recall.}} Endowing intelligent agents with more sophisticated memory systems (e.g., episodic memories) and connecting larger and larger parts of a city to sensors that provide continuous input to AI models can lead to a pervasive 
urban AI that has perfect recall of almost everything that has happened.
We need to understand how we can appropriately limit such systems and endow them with a social AI
that by its design will reliably evolve in harmony with human society.
This might require to endow AI systems with some ethical understanding or memories with the ability to forget.

\paragraph{\textbf{Virtual accessibility of ``everything".}} 
To a large degree the significance of humans, experiences and things for us is related to the ways in which they are special to us. What will happen if sophisticated urban models reach a level of detail and realism that they can begin to compete with the real city and its citizens? How can we develop our ethics to provide guidance in future ``hybrid cities" that not only mix cultures but
also the real and the virtual in unprecedented seamlessness and variability of the virtual?

Thus, urban planning, AI and ethics need rich interactions that deserve careful attention to properly enhance cities as places that foster human potential and creativity.

\section{Conclusions}
Tracing back city models to their historical roots, we have shown their steady development from simple map-like
descriptions to modern digital counterparts that combine many sophisticated methods 
to generate large scale renderings that can represent the physical substrate of cities to a high degree of realism.
We then have set out to analyze what still is largely missing from these approaches: to adequately include the
behavior of the citizens as the most important factor that an urban model must address.
Linking this gap with the recent availability of very large-scale generative models with billions of parameters,
particularly in the form of large language models (LLMs) and their recent multimodal variants,
we have argued that we now are for the first time in a position 
to create rich citizen models that can capture a realistic level of complexity. 
We discussed strengths and weaknesses of these models, particularly their large resource footprints,
the still existing challenges that come with adapting them in a fashion that is reliable and interpretable,
and encouraging developments, particularly the advent of resource-parsimonious small language models
(SLMs) that can approximate the capabilities of LLMs in much smaller footprints. We have then made a
proposal how to bring these different strands of developments together to achieve a new level of urban modeling
where models are interactive and can reach beyond traditional representations of a physical city populated by
rather simple agents that are limited to an almost particle-like behavior due to their small number of parameters.
Instead we have sketched a concept how we can exploit the capabilities of multimodal SLMs to reach into a
modeling of the mental spaces that are contributed by the citizens and that we argue as being at the root
of driving urban development. We have linked this to the concept of a ``Social AI" that brings the citizens' anchoring
in a social world into its focus, along with important driving forces such as resources, desires, needs and sentiments.
We finally came up with a sketch of an architecture that can provide the necessary substrate for such
an endeavor. It achieves resource parsimony by employing a distributed network of nodes each of which combines
SLMs and Digital Twins to effectively combine the holistic reasoning power and human-friendly interaction formats
of the former with the hallucination-free reliability and inspectability of the latter. We argue that this
approach can provide a solid technological base for realizing safe, manageable and resource parsimonious
intelligent ecosystems for driving a new generation of urban models based on Social AI.
We believe that this Social AI can foster not only the development
of the physical spaces of the city but also
the development of its social space which arises
from the interaction of the many mental spaces of the citizens. While this certainly is still a distant vision, innovation from coupling urban models, Digital Twins and generative AI  systems together can already now be seen to bring us ever closer to realising this exciting vision.
 
\backmatter

\bibliography{SAIN}

\begin{appendices}

\section{ChatGPT-4o outputs}

\begin{figure}[t]
{ \caption{The ChatGPT-4o output  on 4 Sept. 2024 given the image and prompt in Fig. \ref{fig:pool-in-wharfedale}.}
    \label{fig:pool-in-wharfedale-output}
    
{The map you provided appears to depict the village of Pool-in-Wharfedale, located near the River Wharfe in the region of West Yorkshire, England. Here's a detailed description of the geographical and built environment features visible on the map:

\#\#\# Geographical Features:
1. **River Wharfe**: A significant geographical feature on the map, the River Wharfe flows from the northwest to the east along the northern edge of the village. The river creates a natural boundary and likely influences the local landscape and activities.

2. **Leathley**: Across the River Wharfe, the area labeled as Leathley is indicated in North Yorkshire. The river serves as a boundary between West Yorkshire (where Pool-in-Wharfedale is located) and North Yorkshire.

3. **Green Spaces**: There are several green areas, particularly to the north and west of the village. These areas are likely open fields, woodlands, or agricultural lands, typical of the rural landscape in this region.

\#\#\# Built Environment Features:
1. **Main Roads**:
   - **A658**: A major road labeled A658 runs through the village from south to north, crossing the River Wharfe via a bridge. This road connects Pool-in-Wharfedale to surrounding towns and regions, making it a key transport route.
   - **A659**: This road intersects with the A658 in the southern part of the village, running from west to east, and continues towards the east as Arthington Lane. 

2. **Residential Areas**:
   - The village is primarily composed of residential areas, with various roads such as Church Close, Mill Lane, and Manor Court serving the local community. These areas are characterized by clusters of houses and small streets.

3. **Community and Religious Buildings**:
   - **Churches**: There are two symbols of churches marked with crosses on the map. These are likely to be significant religious or community buildings.
   - **Pool-in-Wharfedale Church of England Voluntary Controlled Primary School**: Located towards the southeast of the village, this school is an important community institution.

4. **Recreational Areas**:
   - **Pool Cricket Club**: Situated to the northeast of the village, this cricket ground is a notable recreational facility for local residents.
   - The presence of several green spaces and the cricket club indicates recreational opportunities within the village.

5. **War Memorial**: A war memorial is marked in the central part of the village, indicating a place of remembrance and historical significance.

6. **Bridges**: The map shows a bridge where the A658 crosses the River Wharfe, which is a crucial piece of infrastructure connecting different parts of the region.

\#\#\# Surrounding Areas:
- To the north, beyond the River Wharfe, the land appears to be more rural with less dense development, while the village itself is more built-up.

Overall, Pool-in-Wharfedale is depicted as a small village with essential amenities, recreational facilities, and residential areas, set against a backdrop of natural features like the River Wharfe and surrounding green spaces.
}}
   
\end{figure}

\begin{figure}[h]
 \caption{ChatGPT-4o's response on 4/9/2024 to the question ``Please describe a route from the fishmonger to the school.''}
 \label{fig:fishmonger-school-route}
{

   To describe the route from the Fishmonger to the School on the provided map:

1. **Start at the Fishmonger:** 
   - The Fishmonger is located on the right side of the map along the yellow road.

2. **Head left along the Yellow Road:**
   - Move left along the yellow road towards the center of the map. You will pass by the Restaurant, which is centrally located on this road.

3. **Continue past the Newsagent:**
   - Continue traveling left on the yellow road. You will pass the Newsagent, which is located to the left of the Restaurant.

4. **Reach the intersection with the Red Road:**
   - Continue until you reach the intersection where the yellow road crosses the red diagonal road.

5. **Turn left onto the Red Road:**
   - Turn left onto the red road. 

6. **Proceed up the Red Road to the School:**
   - Continue traveling upward along the red road until you reach the School, which is located just slightly below the intersection of the red road and the yellow road.

This route will take you directly from the Fishmonger to the School by primarily using the yellow road and then switching to the red road near the School's location.
  }
\end{figure}

\begin{figure}[h]
  \caption{The response of ChatGPT-4o on 4 Sept 2024 when presented with the sketch map and prompt displayed in Fig. \ref{fig:sketch-map}}
    \label{sketch-map-output}
{
 The sketch map you've provided features a simple layout of a small town or village with various buildings and roads marked. Here’s a detailed description that someone could use to recreate the map:

\#\#\# Roads:
1. **Red Road (Diagonal, Top-Left to Bottom-Right):**
   - Starts from the top left corner and ends at the bottom right corner.
   - Intersects with the green road and the yellow road.

2. **Green Road (Diagonal, Bottom-Left to Top-Right):**
   - Starts from the bottom left corner and ends at the top right corner.
   - Intersects with the red road and the yellow road.

3. **Yellow Road (Horizontal, Middle):**
   - Runs horizontally through the middle of the map.
   - Intersects with the red and green roads.

4. **Purple Road (Horizontal, Bottom):**
   - Runs horizontally near the bottom of the map.
   - Does not intersect with the red or green roads.

\#\#\# Buildings and Locations:
- **Top-Left (near the Red Road):**  
  - **School:** Positioned slightly below where the red road intersects with the yellow road.
  
- **Top-Right (along the Green Road):**
  - **Post Office:** Located just before the green road reaches the top-right corner.

- **Middle of the Map (on the Yellow Road):**
  - **Restaurant:** Centrally located on the yellow road, close to where it intersects with the green road.
  
- **Middle-Left (on the Yellow Road):**
  - **Newsagent:** Positioned to the left of the restaurant along the yellow road.
  
- **Middle-Right (on the Yellow Road):**
  - **Fishmonger:** Positioned to the right of the restaurant along the yellow road.

- **Bottom-Left (along the Purple Road):**
  - **Bakery:** Located at the far left on the purple road.

- **Bottom-Right (along the Purple Road):**
  - **Butcher:** Located at the far right on the purple road.

- **Middle (slightly off the Green Road):**
  - **Pub:** Positioned very close to the intersection of the green road and the yellow road, just below the yellow road.

- **Bottom-Left (along the Green Road):**
  - **Café:** Located where the green road intersects the purple road, but closer to the left side.

- **Bottom-Right (along the Red Road):**
  - **Charity Shop:** Located at the bottom right, close to where the red road meets the purple road.

\#\#\# Notes:
- The buildings are depicted as small blue rectangles.
- The roads are represented by colored lines: red, green, yellow, and purple.
- The roads intersect at various points, creating a cross-like pattern with the red and green roads.

This detailed description should help someone recreate the map accurately.
  
    }
\end{figure}
\end{appendices}

\end{document}